\documentclass[pra,twocolumn,preprintnumbers,amsmath,amssymb,showpacs,
nofootinbib,floatx]{revtex4}

\usepackage{graphicx,bm,amsmath}
\usepackage{hyperref}
\makeatletter
\def\graphicscale{\twocolumn@sw{0.30}{0.4}}
\def\graphicthreescale{\twocolumn@sw{0.30}{0.4}}

\makeatother

\begin{document}

\title{Equilibrium and nonequilibrium entanglement properties
of 2D and 3D Fermi gases}

\author{Jacopo Nespolo and Ettore Vicari } 
\affiliation{Dipartimento di Fisica
dell'Universit\`a di Pisa and INFN, Pisa, Italy}

\begin{abstract}

We investigate the entanglement properties of the equilibrium and
nonequilibrium quantum dynamics of 2D and 3D Fermi gases, by computing
entanglement entropies of extended space regions, which generally show
multiplicative logarithmic corrections to the leading power-law
behaviors, corresponding to the logarithmic corrections to the area
law.

We consider 2D and 3D Fermi gases of $N$ particles constrained within
a limited space region, for example by a hard-wall trap, at
equilibrium at $T=0$, i.e. in their ground state, and compute the
first few terms of the asymptotic large-$N$ behaviors of entanglement
entropies and particle fluctuations of subsystems with some convenient
geometries, which allow us to significantly extend their computation.
Then, we consider their nonequilibrium dynamics after instantaneously
dropping the hard-wall trap, which allows the gas to expand freely.
We compute the time dependence of the von Neumann entanglement entropy
of space regions around the original trap.  We show that at small time
it is characterized by the relation $S \approx \pi^2 V/3$ with the
particle variance, and multiplicative logarithmic corrections to the
leading power law, i.e. $S \sim t^{1-d}\ln(1/t)$.

\end{abstract}


%
% 05.30.Fk, Fermion systems and electron gas 
% 03.65.Ud, Entanglement and quantum nonlocality
% 67.85.-d, Ultracold gases, trapped gases

\maketitle

% ========================= BODY =========================
%\narrowtext

\section{Introduction}
\label{intro}

The nature of the quantum correlations of many-body systems, and in
particular the entanglement phenomenon, are fundamental physical
issues. They have attracted much theoretical interest in the last few
decades, due to the progress in the experimental activity in atomic
physics, quantum optics and nanoscience, which has provided a great
opportunity to investigate the interplay between quantum and
statistical behaviors in particle systems.  In particular, the great
ability in the manipulation of cold
atoms~\cite{CW-02,Ketterle-02,BDZ-08} allows the realization of
physical systems which are accurately described by theoretical models,
such as dilute atomic Fermi and Bose gases, Hubbard and Bose-Hubbard
models, with different effective spatial dimensions from one to three,
achieving thorough experimental checks of the fundamental paradigma of
the condensed matter physics.  Experiments with cold atoms allow to
investigate the unitary quantum evolution of closed many-body systems,
exploiting their low dissipation rate which maintains phase coherence
for a long time~\cite{BDZ-08,PSSV-11}.  In this experimental context,
the theoretical investigation of nonequilibrium dynamics in quantum
many-body systems, and the time evolution of the entanglement
properties characterizing the quantum correlations, is of great
importance for a deep understanding of the fundamental issues of
quantum dynamics, their possible applications, and new developments.

Quantum correlations are characterized by the fundamental phenomenon
of entanglement, which gives rise to nontrivial connections between
different parts of extended quantum
systems~\cite{CCD-09,AFOV-08,ECP-10}.  A measure of entanglement is
provided by the entanglement entropies associated with the reduced
density matrix $\rho_A={\rm Tr}_B \rho$ of a subsystem $A$ with
respect to its complement $B$.  The entanglement properties of 1D
systems have been substantially understood, and in particular a
leading logarithmic behavior has been established at conformally
invariant quantum critical points~\cite{CC-09}. However, in higher
dimensions the scaling behavior of the bipartite entanglement entropy
is generally more complicated, without a definite general scenario for
the universal behavior at a quantum phase transition.

In general, the entanglement entropy is asymptotically proportional to
the surface separating the two subsystems $A$ and $B$.  The area law
has been generally proven for gapped systems \cite{WVHC-08},
independently of the statistics of the microscopical constituents.  On
the other hand, lattice free fermions show multiplicative logarithmic
corrections to the area law
\cite{Wolf-06,GK-06,BCS-06,LDYRH-06,HLS-09,DBYH-08,Sobolev-10,%%
  Swindle-10,CMV-12}. Indeed, for a large subsystem $A$ of size $l_A$
the entanglement entropy scales as
\begin{equation}
S^{(\alpha)}_A \sim l_A^{d-1}\ln l_A,
\label{carelaw}
\end{equation}
where $S^{(\alpha)}$ are the R\'enyi entropies defined as
\begin{equation}
S^{(\alpha)}_A = \frac{1}{1-\alpha} \ln {\rm Tr}\rho_A^{\alpha},
\label{ren}
\end{equation}
whose limit $\alpha\to 1$ gives the von Neumann (vN) entropy
\begin{equation}
S^{(1)}_A = - {\rm Tr}\rho_A \ln \rho_A.
\label{renvN}
\end{equation}
Analogously, in $d$-dimensional systems of $N$ noninteracting fermions
constrained in a finite volume, the entanglement entropy of a
subsystem $A$ shows multiplicative logarithmic corrections to the
leading power-law behavior. Indeed it grows asymptotically
as~\cite{CMV-12}
\begin{equation}
S^{(\alpha)}_A\sim N^{(d-1)/d} \ln N.
\label{lnlog}
\end{equation}
The logarithm of the large-$N$ asymptotic behavior can be related to
the logarithmic area-law violation (\ref{carelaw}) of lattice free
fermions, by considering the thermodynamic limit $N,L\to\infty$
keeping the particle density $\rho=N/L^d$ fixed~\cite{CMV-12}.  The
leading large-$N$ asymptotic behavior, including its coefficient, can
be computed exploiting the so-called Widom
conjecture~\cite{Widom-81,GK-06,HLS-09,Sobolev-10,CMV-12}.  However,
this approach does not provide information on the next-to-leading
terms. These have been conjectured to be logarithmically suppressed
with respect to the leading term~\cite{CMV-12}, as in the case of 1D
Fermi gases~\cite{CMV-11,CMV-11-2} where there are alternative and
more powerful approaches to compute the asymptotic behavior of the
entanglement entropy, based on the Fisher-Hartwig
conjecture~\cite{FH-68,BM-94} and generalizations~\cite{idk-09}.

In this paper we further investigate this issue. We consider Fermi
gases of $N$ particles in a finite $L^d$ volume at equilibrium at
$T=0$, i.e. in their ground state, and compute the first few terms of
the asymptotic large-$N$ behaviors of entanglement entropies and
particle fluctuations of subsystems with some convenient geometries,
which allow us to significantly extend their computation.  For
example, the half-space vN entanglement entropy of a system of size
$L^d$ turns out to asymptotically increase as (setting $L=1$)
\begin{equation}
S^{(1)}_{\rm HS} =  a N^{(d-1)/d} \left[ \ln N + a_0 + 
O\left(N^{-1/d}\right) \right].
\label{asym}
\end{equation}
We exactly compute the constants $a$ and $a_0$ for 2D and 3D, for both
periodic boundary conditions (PBC) and open (hard-wall) boundary
conditions (OBC).  The calculation of coefficient $a$ confirms the
results already obtained using the Widom conjecture, while that of the
constant $a_0$ provides new information. Moreover, we investigate the
successive terms of the expansion which are generally suppressed by
powers of $1/N$, and show peculiar oscillations.

Beside the ground-state properties, we consider the nonequilibrium
dynamics of Fermi gases after dropping the hard walls of the trap,
which allows the gas to expand freely.  The free expansion of gases
after the drop of the trap is routinely exploited in experiments to
infer the properties of the initial quantum state of the particles
within the trap~\cite{BDZ-08}, by observing the interference patterns of
absorption images in the large-time ballistic regime.  The quantum
dynamics of freely expanding fermionic systems has been much
investigated, see e.g.
Refs.~\cite{BDZ-08,CGM-09,OS-02,PSOS-03,MG-05,RM-05,CM-06,GP-08,%%
Gerbier-etal-08,HRMFD-08,HMRMFD-09,LSMSH-12,V-12-2,BHLMOR-12,Schneider-etal-12},
considering also particle systems constrained into lattice structures.
In particular, Ref.~\cite{Schneider-etal-12} reports an experimental
investigation of the trasport properties during the expansion of
fermionic atoms in optical lattices.

Beside being of phenomenological interest, a Fermi gas released from a
trap is one of the simplest examples of nonequilibrium unitary
evolution, which may provide interesting general information on the
nonequilibrium entanglement properties.  We are interested in the time
dependence of extended observables around the original hard-wall trap,
such as entanglement entropies and particle fluctuations associated
with finite space regions, during the nonequilibrium dynamics after
the drop of the trap.  This issue was already studied in 1D Fermi
gases~\cite{V-12-2}, finding a logarithmic small-time behavior,
$S^{(1)}\approx (1/3)\ln(1/t)$, for the vN entanglement entropy of
intervals in proximity to the initial trap, and the relation
\begin{equation}
S^{(1)}\approx \frac{\pi^2}{3} V^{(2)}
\label{s1vint}
\end{equation}
with the particle variance $V^{(2)}$. These results resemble equilibrium
behaviors~\cite{CC-04,CMV-11,CMV-12-2} where the leading logarithms
are essentially determined by the corresponding conformal field theory
with central charge $c=1$. We show that analogous relations hold in
the nonequilibrium dynamics of higher-dimensional Fermi gas released
from hard-wall traps.  In the large-$N$ regime we exactly determine
the small-$t$ behavior of the entanglement entropy of spatial
subsystems $A$ in proximity of the original trap. We find that the
relation (\ref{s1vint}) still holds, and that its asymptotic 
small-$t$ behavior,
\begin{equation}
S^{(1)}_A(t) \sim t^{1-d} \ln(1/t),
\label{s1ati}
\end{equation}
is characterized by a multiplicative logarithmic correction to the
leading power law.

The paper is organized as follows.  In Sec.~\ref{mbwaf} we report the
many-body wave function describing the ground state of $N$-particle
Fermi gases in cubic-like hard-wall traps, and their free expansion
after dropping the trap; moreover we define the extended observables
that we consider.  In Sec.~\ref{strip} we compute the asymptotic
large-$N$ behavior of the entanglement entropies and particle
fluctuations of strip-like subsystems, which allow us to exactly
compute the leading and next-to-leading terms and investigate the
power-law suppressed (oscillating) corrections. In Sec.~\ref{sphvol}
we consider Fermi gases in a disk, to investigate the dependence of
the large-$N$ asymptotic behavior on the system's shape.
Section~\ref{reltrap} is devoted to the study of the entanglement
properties of the nonequilibrium dynamics of $d$-dimensional
noninteracting Fermi gases released from a trap; in particular, we
determine its small-$t$ and large-$t$ asymptotic behaviors for
subsystems corresponding to the original trap.  Finally, in
Sec.~\ref{conclusions} we summarize our main results and draw our
conclusions.

\section{Many-body wave functions and extended observables}
\label{mbwaf}

\subsection{Fermi gas within hard walls}
\label{grwf}

We consider a $d$-dimensional gas of $N$ spinless noninteracting
fermionic particles of mass $m$, confined within a limited space
region by a hard-wall trap. For simplicity, we mostly consider
cubic-like traps of size $L$, and place the origin of the axes at the
center of the trap, so that the allowed region for the particles is
\begin{equation}
T=[-L/2,L/2]^d.
\label{trapreg}
\end{equation}  
In the following we set $\hslash=1$ and $m=1$.

The ground-state wave function is
\begin{equation}
\Psi({\bf x}_1,...,{\bf x}_N) = {1\over \sqrt{N!}} {\rm det} 
[\psi_n({\bf x}_m)],
\label{fpsi}
\end{equation}
where $\psi_n$ are the lowest $N$ eigensolutions of the free
one-particle Schr\"odinger equation with the appropriate boundary
conditions. For example, assuming open (hard-wall) boundary
conditions, the eigensolutions can be written as a product of
eigenfunctions of the corresponding 1D Schr\"odinger problem, i.e.
\begin{eqnarray}
&&\psi_n({\bf x}) \equiv \psi_{n_1,n_2,...,n_d}({\bf x})= 
\prod_{i=1}^d \varphi_{n_i}(x_i),
\label{prodfunc}\\
&&E_{n_1,n_2,...,n_d}= \sum_{i=1}^d e_{n_i},
\label{sunei}
\end{eqnarray}
where the subscript $n_i$ labels the eigenfunctions along the $d$
directions, which satisfy $\varphi_n(-L/2)=\varphi_n(L/2)=0$, thus
\begin{eqnarray}
\varphi_n(x) = {1\over \sqrt{L/2}} {\rm sin}\left(n \pi 
{x+L/2\over L}\right),\quad
e_{n} =  {\pi^2\over 2L^2} n^2, \label{1deigf}
\end{eqnarray}
for $n=1,2,...$. 

The one-particle correlation function reads
\begin{eqnarray}
C({\bf x},{\bf y}) \equiv  
\langle c^\dagger({\bf x}) 
c({\bf y}) \rangle = 
\sum_{n=1}^N \psi_n({\bf x})^*\psi_n({\bf y}),
 \label{rhonbos}
\end{eqnarray}
where $c({\bf x})$ is the fermionic annihilation operator. The
particle density $\rho({\bf x})=\langle n({\bf x})\rangle$, where
$n({\bf x})\equiv c^\dagger({\bf x}) c({\bf x})$, and the connected
density-density correlation $G_n({\bf x},{\bf y}) \equiv \langle
n({\bf x}) n(\bf{y})\rangle_c$ can be easily derived from the
two-point function; for example $\rho({\bf x})= \sum_{n=1}^N
|\psi_n({\bf x})|^2$.

Other interesting observables are related to the particle-number
distribution over a spatial region $A$, which is described by the
expectation value and correlations of the particle-number operator
$\hat{N}_A\equiv \int_A d^d x\, n({\bf x})$.  A more powerful
characterization of the particle distribution is achieved using its
cumulants $V_A^{(m)}$ ~\cite{cumgen}.  In particular, the particle
variance reads
\begin{equation}
V_A^{(2)}=
\langle {\hat N}_A^2 \rangle -\langle {\hat N}_A \rangle^2.
\label{v2na}
\end{equation}
A measure of the entanglement of the extended region $A$ with the rest
of the system is provided by the vN and R\'enyi entanglement
entropies, cf. Eq.~(\ref{ren}).

In noninteracting Fermi gases the particle cumulants and the
entanglement entropies of a subsystem $A$ can be related to the
two-point function $C$ restricted within $A$~\cite{Peschel-03}, which
we denote by $C_A({\bf x},{\bf y})$.  For example, the particle number
and cumulants within $A$ can be derived using the relations $N_A =
{\rm Tr}\,C_A$ and~\cite{SRFKLL-12}
\begin{eqnarray}
&& V_A^{(m)} = (-i\partial_z)^m {\cal G}(z,C_A)|_{z=0},\label{vnyc}\\
&&{\cal G}(z,{\mathbb X}) = {\rm Tr}\ln\left[1 + \left(e^{iz} - 1\right)
{\mathbb X}\right].
\label{ygenc}
\end{eqnarray}

The computation of the particle cumulants and entanglement entropies
in Fermi gases of $N$ particles is much simplified by introducing the
$N\times N$ overlap matrix ${\mathbb A}$,
\begin{equation}
{\mathbb A}_{nm} =  \int_A d^d x\, \psi_n^*({\bf x}) \psi_m({\bf x}),
\qquad n,m=1,...,N,
\label{aiodef}
\end{equation}
where the integration is over the spatial region $A$, and involves the
lowest $N$ energy levels. Indeed, exploiting the relation
\begin{equation}
{\rm Tr}\,{\mathbb A}^k = {\rm Tr}\,C_{A}^k 
\label{acrel}
\end{equation}
for any $k$, one can compute the particle cumulants and the
entanglement entropies from the eigenvalues $a_n$ of the overlap
matrix~\cite{CMV-11,CMV-11-2,CMV-12-2,V-12}. For example
\begin{eqnarray}
S^{(\alpha)}_A = {1\over 1-\alpha}  \sum_{n=1}^N 
\ln \left[{a_n}^\alpha
+\left({1-a_n}\right)^\alpha\right].
\label{soma}
\end{eqnarray}
In the limit $\alpha\to 1$ we recover the formula for the vN
definition.

For noninteracting fermions, one can write down a formal expansion of
the entanglement entropies of bipartitions in terms of the even
cumulants $V^{(2k)}_A$ of the particle-number distribution. Indeed,
the R\'enyi entropies can be written as~\cite{KL-09,SRFKL-11}
\begin{eqnarray} 
&& S^{(\alpha)}_A= \sum_{k=1}^\infty s^{(\alpha)}_k V^{(2k)}_A,
\label{sv2ia} \\
&& s^{(\alpha)}_k= 
{(-1)^k (2\pi)^{2k}2  \zeta[-2k,(1+\alpha)/2]
\over [ (\alpha-1) \alpha^{2k} (2k)!]}, 
\nonumber
\end{eqnarray} 
where $\zeta$ is the generalized Riemann zeta function.  The expansion
(\ref{sv2ia}) gets effectively truncated in the large-$N$ limit.
Indeed, in Eq.~(\ref{sv2ia}) the leading $O(N^{(d-1)/d} \ln N)$
asymptotic behavior of $S^{(\alpha)}_A$ arises from $V^{(2)}_A$ only,
because the leading order of each cumulant $V^{(m)}$ with $m>2$
vanishes for any subsystem $A$ (including disjoint ones) in any
dimension~\cite{CMV-12-2}, and also in the presence of an external
space-dependent confining potential~\cite{V-12}.  This implies the
general asymptotic relation~\cite{CMV-12-2}
\begin{equation} 
{S^{(\alpha)}_A/V^{(2)}_A} =
{(1+\alpha^{-1})\pi^2/6} + o(1). \label{anyd} 
\end{equation}

\subsection{Fermi gas released from the hard-wall trap}
\label{freexp}

We also consider the nonequilibrium evolution of a Fermi gas initially
in its ground state within the hard-wall trap (\ref{trapreg}), after
the instantaneous drop of the trap.  This is described by the
time-dependent wave function
\begin{equation}
\Phi({\bf x}_1,...,{\bf x}_N;t) 
= {1\over \sqrt{N!}} {\rm det}  [\phi_n({\bf x}_m,t)]
\label{fpsit}
\end{equation}
where $\phi_n({\bf x},t)$ are the one-particle wave functions with
initial condition $\phi_n({\bf x},0)=\psi_n({\bf x})$, corresponding
to the lowest $N$ states of the one-particle Hamiltonian at
$t=0$. They can be written using the free propagator ${\cal P}$ as
\begin{eqnarray}
&&\phi_n({\bf x},t) = \int_T d^dy 
\,{\cal P}({\bf x},t;{\bf y},0) \psi_n({\bf y}),\label{fexev}\\
&&{\cal P}({\bf x},t;{\bf y},0) =
\prod_{i=1}^d P(x_i,t;y_i,0),  \nonumber\\
&&P(x,t;y,0)= {1\over \sqrt{i 2\pi t} } 
\exp\left[{i(x-y)^2\over 2 t}\right].
\nonumber
\end{eqnarray}

The equal-time observables can again be written in terms of the
one-particle wave functions, essentially by replacing the equilibrium
wave functions $\psi_n({\bf x})$ with the time-dependent wave function
$\phi_n({\bf x},t)$. For example, the equal-time one-particle and
connected density-density correlation functions read
\begin{eqnarray}
&&C({\bf x},{\bf y},t)
=\sum_{n=1}^N \phi_n({\bf x},t)^*\phi_n({\bf y},t),
\label{tpf}\\
&&G_n({\bf x},{\bf y},t)
=-|C({\bf x},{\bf y},t)|^2 + \delta({\bf x}-{\bf y}) 
C({\bf x},{\bf y},t).\qquad \label{gnbost}
\end{eqnarray}
The particle cumulants and the entanglement entropies of a subsystem
$A$ can be again related to the two-point function $C$ restricted
within $A$~\cite{PE-09}. One can also define a time-dependent overlap
matrix~\cite{V-12-2}
\begin{equation}
{\mathbb A}_{nm}(t) =  \int_A d^dx\, \phi_n^*({\bf x},t) \phi_m({\bf x},t),
\qquad n,m=1,...,N,
\label{aiodeft}
\end{equation}
where the integration is over the spatial region $A$, and involves the
time evolutions $\phi_n({\bf x},t)$ of the lowest $N$ energy states of
the one-particle Schr\"odinger problem at $t=0$.  The time-dependent
particle fluctuations within $A$ and the bipartite entanglement
entropies can be again computed form the time-dependent eigenvalues of
${\mathbb A}$, using formulas analogous to those at
equilibrium~\cite{V-12-2}. In particular, also Eq.~(\ref{sv2ia}) holds
during the time evolution.

\section{Entanglement entropy of strip-like subsystems in 2D and 3D}
\label{strip}

\subsection{Overlap matrix for strip-like subsystems}
\label{omstrip}

\begin{figure}[t]
 \begin{center}
 \includegraphics[width=4.5cm,keepaspectratio=true]{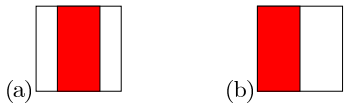}
 \end{center}
 \caption{(Color online) A schematic representation of the subsystems
   (shadowed) considered in 2D: (a) a strip centered within the
   system; (b) a strip starting at one edge.  Of course, in the case
   of PBC they are equivalent.  }
 \label{fig:subsys_diagram}
\end{figure}

We consider Fermi gases constrained within the trap region
(\ref{trapreg}) with open (hard-wall) boundary conditions (OBC) and
periodic boundary conditions (PBC). The leading large-$N$ behavior
(\ref{lnlog}) of the entanglement entropy of any connected subsystem
has been computed by applying the Widom conjecture to evaluate the
asymptotic behavior of the formal expressions derived from the overlap
matrix.  However, this approach does not provide any information on
the next-to-leading terms.

We now extend these calculations for some convenient strip-like
geometries
\begin{eqnarray}
&&A = {\cal I}\times [-L/2,L/2]^{d-1},\label{Astrip}\\
&& {\cal I} = [-\delta L/2,\delta L/2],  \qquad 0<\delta<1,
\label{calA}
\end{eqnarray}
see Fig.~\ref{fig:subsys_diagram}a. This choice of subsystems
significantly simplifies the computations, allowing
us to exactly derive the leading and next-to-leading terms of the
large-$N$ expansion, and to identify the power law of the corrections.

For our purpose, we exploit the fact that the corresponding overlap
matrix (\ref{aiodef}) is a block diagonal matrix. Indeed, relabeling
the indexes $n,m$ of the $N\times N$ overlap matrix as $n_1,...,n_d$
and $m_1,...,m_d$, and using Eq.~(\ref{prodfunc}), we can write the
half-space overlap matrix as
\begin{eqnarray}
&&{\mathbb A}_{n_1,...,n_d;m_1,...,m_d} =
{\mathbb E}_{n_1m_1} \prod_{i=2}^d \delta_{n_im_i} ,
\label{ahs}\\
&&{\mathbb E}_{nm} =
\int_{\cal I} dx\, \varphi_{n}(x) \varphi_{m}(x),
\label{ahs1d}
\end{eqnarray}
$\varphi_{n}$ are the 1D eigenfunctions on the interval ${\cal I}$
(for example those reported in Eq.~(\ref{1deigf}) for OBC), the
indexes $n_1,...,n_d$ correspond to the lowest $N$ states according to
Eq.~(\ref{sunei}).

\subsection{Leading and next-to-leading terms in 2D systems}
\label{2dsy}

\subsubsection{Periodic boundary conditions}
\label{pbc2d}

Let us first consider a 2D system with PBC.  The corresponding
one-particle eigenspectrum is given by Eqs.~(\ref{prodfunc}) and
(\ref{sunei}) with
\begin{eqnarray}
\varphi_n(x) = {1\over \sqrt{L}} e^{i 2 \pi n (x+L/2)/L}, 
\quad e_{n} = {2 \pi^2\over L^2} n^2, \label{pbcs}
\end{eqnarray}
for $n\in {\mathbb Z}$.  
We construct the ground state of a Fermi
gas by filling all states with
\begin{equation}
n_1^2+n_2^2\le n_f^2,\qquad n_i \in {\mathbb Z}, \label{nedef}
\end{equation}
and $n_f$ is a real number.  The number $N$ of particles is a function
of $n_f$, which asymptotically reads $N\approx \pi n_f^2$.  In the
following we set $L=1$.

Since the overlap matrix (\ref{ahs}) is block diagonal, for any
integer $k$ we have
\begin{eqnarray}
{\rm Tr}\,{\mathbb A}[N(n_f)]^k = 
\sum_{n_1=-{\lfloor n_f \rfloor}}^{\lfloor n_f \rfloor} 
{\rm Tr}\,{\mathbb E}(2\lfloor \sqrt{n_f^2-n_1^2} \rfloor + 1)^k , 
\label{adetrap}
\end{eqnarray}
where $\lfloor x \rfloor$ indicates the largest integer which is
smaller than $x$, and ${\mathbb E}(M)$ is the $M\times M$ overlap
matrix (\ref{ahs1d}) constructed using the lowest $M$ 1D states.  This
also implies analogous exact relations for all observables $O_A$ which
are functions of the eigenvalues of the overlap matrix, such as the
particle cumulants and the entanglement entropies. Thus,
\begin{eqnarray}
O_{A} = 
\sum_{n_1=-\lfloor n_f \rfloor }^{\lfloor n_f \rfloor} 
{\cal O}_{\cal I}\left(2\lfloor
 \sqrt{n_f^2-n_1^2} \rfloor + 1\right),
\label{oatrap}
\end{eqnarray}
where ${\cal O}_{\cal I}(M)$ is
the corresponding quantity for 1D $M$-particle
systems.

In order to derive their large-$N$ asymptotic behaviors, we replace
the sum by an integral
\begin{eqnarray}
O_{A}(N) = 
\int_{-n_f}^{n_f} dn \;{\cal O}_{\cal I}\left(2\sqrt{n_f^2-n^2}\right),
\label{sadetrapi}
\end{eqnarray}
where $n_f$ is related to the particle number by
the asymptotic relation $N=\pi n_f^2$.

Then, we derive the asymptotic large-$N$ behavior of the entanglement
entropies and the particle cumulants by inserting the corresponding
asymptotic formulas for the 1D quantities, which have been computed
using the Fisher-Hartwig conjecture~\cite{FH-68,BM-94} and
generalizations~\cite{idk-09,FC-11}.  They are~\cite{CMV-11,CMV-11-2}
\begin{equation}
{\cal S}^{({\alpha})}_{\cal I}(M)  = c_\alpha
\left[ \ln M + \ln\sin(\pi\delta) + b_\alpha \right] + O(M^{-2/\alpha}),
\label{FHres2pbc}
\end{equation}
where 
\begin{eqnarray}
c_\alpha &=& {1+\alpha^{-1}\over 6},  \label{calpha}\\
b_\alpha &=& \ln 2 + \int_0^\infty {dt\over t} 
\times \label{ba}\\
&&\times
\Bigl[{6\over 1-\alpha^{-2}}
\Bigl({1\over \alpha\sinh t/\alpha}  -{1\over\sinh t}\Bigr)
{1\over\sinh t}- e^{-2t}\Bigr].\nonumber
\end{eqnarray}
Concerning the particle cumulants, we have~\cite{CMV-12-2} 
\begin{eqnarray}
&&{\cal V}^{(2)}_{\cal I} = 
{1\over \pi^2}\left[ \ln M+ \ln\sin(\pi\delta)+v_2\right] + O(M^{-2}),\quad
\label{Vres2}\\
&&{\cal V}^{(2k)}_{\cal I} = v_{2k} + O(M^{-\varepsilon})
\quad {\rm for}\;\;k>1,\label{Vres4}
\end{eqnarray}
where 
\begin{eqnarray}
&&v_2=1+\gamma_E + \ln 2,
\label{v21d}\\
&&v_4=-0.0185104,\quad v_6=0.00808937, 
\label{v461d}
\end{eqnarray}
 etc....  The odd cumulants are suppressed, 
\begin{equation}
{\cal V}^{(2k+1)}_{\cal I} = O(M^{-\varepsilon}) \qquad (k\ge 1).  
\label{vodd1d}
\end{equation}
The power $\varepsilon>0$ of the $O(M^{-\varepsilon})$ corrections in
the particle cumulants ${\cal V}^{(m)}$ with $m\ge 3$ is not known.
Numerical results suggest that $\varepsilon$ depends on $m$ and that
it gets smaller with increasing $m$. Note that, since the interval
${\cal I}$ has two boundaries, analogous formulas hold in the case of
OBC, with the only difference that corrections are $O(M^{-1/\alpha})$
in Eq.~(\ref{FHres2pbc}), and $O(M^{-1})$ in Eq.~(\ref{Vres2}).

Inserting the above asymptotic 1D formulas into Eq.~(\ref{sadetrapi}),
 we obtain the desired 2D results.  The asymptotic behavior of the
 entanglement entropies turns out to be
\begin{eqnarray}
&&S_{A}^{(\alpha)} \approx  S_{A,\rm asy}^{(\alpha)} + O(N^{1/2-\kappa}), 
\label{sadetrapr}\\
&&S_{A,\rm asy}^{(\alpha)} =
{c_\alpha\over \sqrt{\pi}}  N^{1/2}\left(\ln N + a_0 \right),\label{asysa}\\
&&a_0 = 2\ln\sin\pi\delta + 2b_\alpha - \ln\pi + 4\ln 2 - 2.
\label{a0asysa}
\end{eqnarray}
The approximations to derive this asymptotic behavior from
Eq.~(\ref{oatrap}) do not affect the leading $O(N^{1/2}\ln N)$ and
next-to-leading $O(N^{1/2})$ terms, so that the constants $a$ and
$a_0$ should be considered as exact.  Indeed, in the derivation of
Eq.~(\ref{sadetrapi}), replacing the sum with an integral induces
$O(N^{-1/2})$ relative errors. The asymptotic 1D behaviors which we
inserted in the integrals for the entanglement entropy lead to
$O(n^{-2/\alpha})$, thus $O(N^{-1/\alpha})$, corrections; while the
fact that we use it also when $n$ is not large should cause
$O(N^{-1/2})$ relative corrections (apart from logarithms).  This
implies that the overall corrections for the entanglement entropies
are those reported in Eq.~(\ref{sadetrapr}) with
\begin{equation}
\kappa={\rm min}[1/2,1/\alpha].
\label{sccorrspbc}
\end{equation}
This will be further checked by numerical calculations below.

For the particle cumulants we obtain
\begin{eqnarray}
&&V_{A}^{(2)} = V_{A,\rm asy}^{(2)}+O(N^0),
\label{v2adetrapr}\\
&&V_{A,\rm asy}^{(2)}= {1\over \pi^{5/2}}
N^{1/2}\left( \ln N + w_0 \right),\label{asyva}\\
&&w_0 = 2\ln\sin\pi\delta + 2v_2 - \ln\pi + 4\ln 2 - 2,
\label{w0asyva}
\end{eqnarray}
and
\begin{eqnarray}
V_{A}^{(m)} \approx 
{2v_m\over \sqrt{\pi}} \,N^{1/2} ,\quad m>2,
\label{vmadetrapr}
\end{eqnarray}
where $v_m$ are the constants of the leading large-$N$ behavior in 1D,
cf. Eqs.~(\ref{v21d}), (\ref{v461d}) and (\ref{vodd1d}).  In the case
of the particle variance the corrections are expected to be suppressed
by $N^{-1/2}$ with respect to the leading term.

Of course, the leading terms with the multiplicative logarithms are in
agreement with the results obtained by the Widom
conjecture~\cite{CMV-12,CMV-12-2}.

\subsubsection{Open boundary conditions}
\label{obc2d}

In the case of OBC the ground state of the Fermi gas is obtained by
filling all states with
\begin{equation}
n_1^2+n_2^2\le n_f^2,\qquad n_i \in {\mathbb N}, \label{nedefo}
\end{equation}
where $n_f$ is asymptotically related to the particle number
by $N\approx (\pi/4) n_f^2$.  
We have again the general relations
\begin{eqnarray}
{\rm Tr}\,{\mathbb A}[N(n_f)]^k = 
\sum_{n_1=1}^{\lfloor n_f\rfloor} {\rm Tr}\,{\mathbb E}
\left(\lfloor \sqrt{n_f^2-n_1^2} \rfloor\right)^k 
\label{adetrapobc}
\end{eqnarray}
for any $k\in {\mathbb N}$, between the trace of powers of the
corresponding overlap matrix ${\mathbb A}$ and the reduced 1D 
${\mathbb E}$, cf. Eq.~(\ref{ahs}).  Then, in the case of the
subsystem $A$, cf. Eq.~(\ref{Astrip}), one can easily check that the
same asymptotic behaviors of PBC apply.  One should however expect
that corrections to the above asymptotic formulas are larger,
essentially because the 1D asymptotic behaviors have larger
corrections. Indeed, for the entanglement entropies we expect that
they are $O(N^{-\kappa})$ with $\kappa=1/(2\alpha)$ with respect to
the leading term.  The particle variance should have $O(N^{-1/2})$
corrections as well.

The above results change if we consider strip-like subsystems which
start from the boundary, essentially because the corresponding
asymptotic 1D behaviors change~\cite{CMV-11,CMV-12-2}.  For example, we
apply the same approach to the half-space subsystem defined as
\begin{eqnarray}
H = [-L/2,0]\times [-L/2,L/2]^{d-1},\label{Astript}
\end{eqnarray}
and shown in Fig.~\ref{fig:subsys_diagram}b.  In this case the
corresponding 1D asymptotic large-$N$ behaviors change~\cite{CMV-11,CMV-12-2},
for example
\begin{equation}
{\cal S}^{({\alpha})}_{\cal H}(M)  = {c_\alpha\over 2}
\left[ \ln M + \ln 2 + b_\alpha \right] + O(M^{-1/\alpha})
\label{FHres2obc}
\end{equation}
where ${\cal H}=[-L/2,0]$, essentially because only one boundary
separates the two parts.  The asymptotic behaviors of the half-space
entanglement entropies and particle cumulants are
\begin{eqnarray}
&&S_H^{(\alpha)} = S_{H, {\rm asy}}^{(\alpha)} + O(N^{1/2-1/(2\alpha)}),
\label{asysahs}\\
&&S_{H,{\rm asy}}^{(\alpha)} =
{c_\alpha\over 2\sqrt{\pi}} N^{1/2}( \ln N + a_0 ),\label{asysahs2}\\
&&a_0 =  2b_\alpha - \ln\pi + 6\ln 2 - 2,
\label{a0asysahs}
\end{eqnarray}
and
\begin{eqnarray}
&&V_{H}^{(2)} = V_{H, {\rm asy}}^{(2)} + O(N^{0})
\label{asyv2hs}\\
&&V_{H,{\rm asy}}^{(2)} =
{1\over 2\pi^{5/2}} N^{1/2}( \ln N + w_0 ),\label{asyv2hs2}\\
&&w_0 = 2v_2 - \ln\pi + 6\ln 2 - 2,
\label{a0asyv2hs}
\end{eqnarray}
\begin{eqnarray}
V_{H}^{(m)} \approx 
{v_m\over\sqrt{\pi}} \,N^{1/2} ,\quad m>2.
\label{vmobc}
\end{eqnarray}

\subsection{Leading and next-to-leading terms 
of the entanglement entropy in 3D systems}
\label{3dsy}

The above calculations can be easily extended to strip-like subsystems
in higher-dimensional systems.  We construct the ground state 3D Fermi
gases with PBC, by filling all states with
\begin{equation}
n_1^2+n_2^2+n_3^2\le n_f^2,\qquad n_i \in {\mathbb Z}, \label{nedef3d}
\end{equation}
and the number $N$ of particles is a function of $n_f$,
asymptotically $N=4\pi n_f^3/3$. 
Exploiting again the block structure of the corresponding overlap matrix
(\ref{ahs}), we can write any observable $O_A$, defined in terms of
the eigenvalues of $A$, as
\begin{eqnarray}
O_{A} = 4 \int_0^{n_f} dn_1 \int_0^{\sqrt{n_f^2-n_1^2}} dn_2 \,
{\cal O}_{\cal I}\left(2\sqrt{n_f^2-n_1^2-n_2^2}\right).
\nonumber\\
\label{intappoa3d}
\end{eqnarray}
Again, by replacing the large-$N$ behavior ${\cal S}_{\cal I}$ of the 1D, we
obtain
\begin{eqnarray}
&& S_A^{(\alpha)} =  S_{A,\rm asy}^{(\alpha)} 
+O(N^{2/3-\kappa}),\label{saasy3d}\\
&&S_{A,\rm asy}^{(\alpha)}=
{\pi^{1/3} c_\alpha\over 2^{4/3}3^{1/3}} N^{2/3}\left(\ln N + a_0\right),
\label{sa3D}\\
&& a_0 = 3\ln\sin\pi\delta + 3b_\alpha - \ln\pi + \ln 6 - 3/2.
\nonumber
\end{eqnarray}
The same asymptotic behaviors apply to the case of OBC.  An
analysis of the errors induced by the approximations indicate that the
corrections are suppressed by $O(N^{-\kappa})$, with $\kappa={\rm
  min}[1/3,2/(3\alpha)]$ for PBC
and $\kappa=1/(3\alpha)$ for OBC, with respect to the leading term.
Analogous results can be obtained for the particle fluctuations.

In the case of OBC we also consider the half-space region
(\ref{Astript}), obtaining
\begin{eqnarray}
&&S_{H}^{(\alpha)} = S_{H,{\rm asy}}^{(\alpha)} + O(N^{2/3-1/(3\alpha)}),
\label{asysahs3}\\
&&S_{H,{\rm asy}}^{(\alpha)} =
{\pi^{1/3} c_\alpha\over 2^{7/3}3^{1/3}}
N^{2/3}( \ln N + a_0 ),\label{asysahs33}\\
&&a_0 = 3b_\alpha - \ln\pi + \ln 24 - 3/2.
\label{a0asysahs3}
\end{eqnarray}

\subsection{Oscillating corrections to the asymptotic behavior}
\label{corrections}

\begin{figure}[tbp]
\includegraphics[width=8.5cm,keepaspectratio=true]{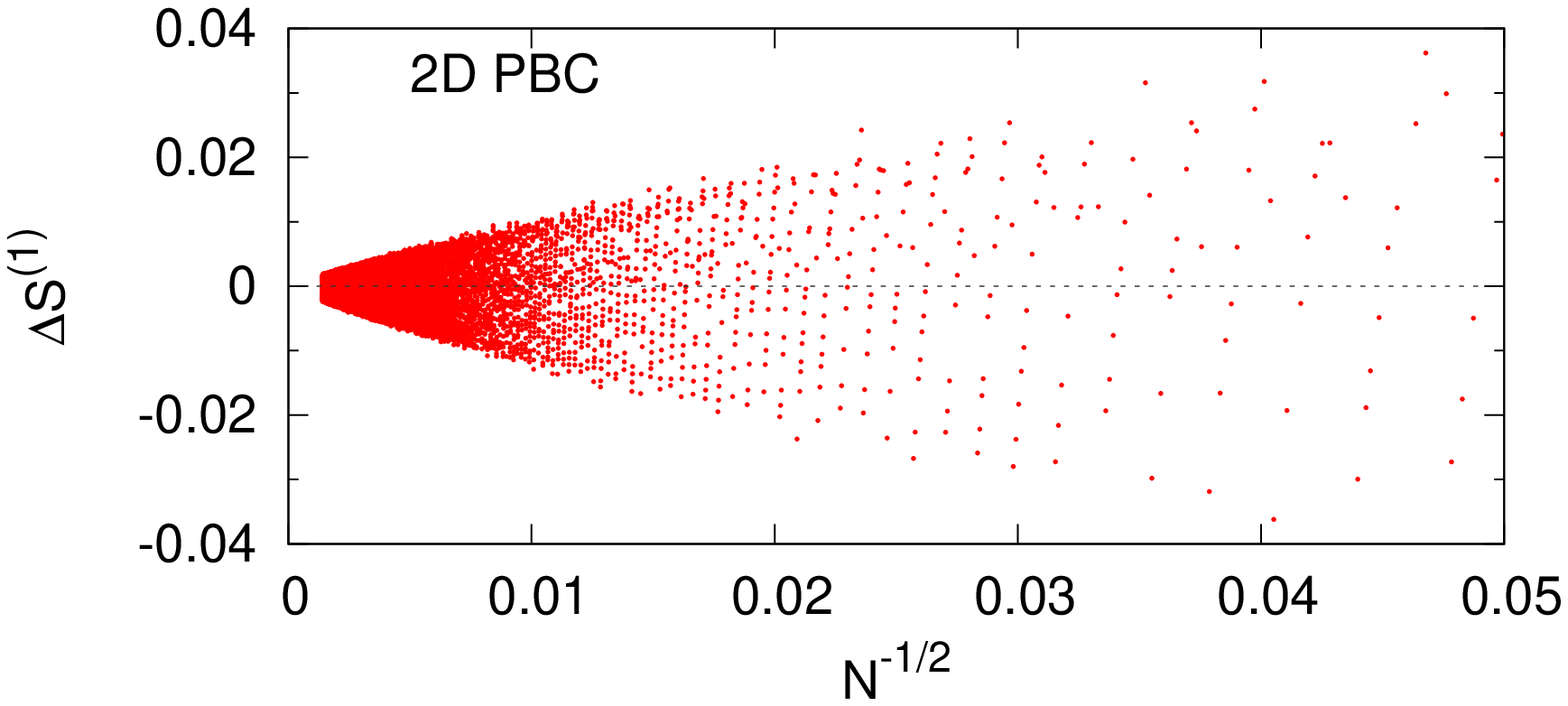}
\includegraphics[width=8.5cm,keepaspectratio=true]{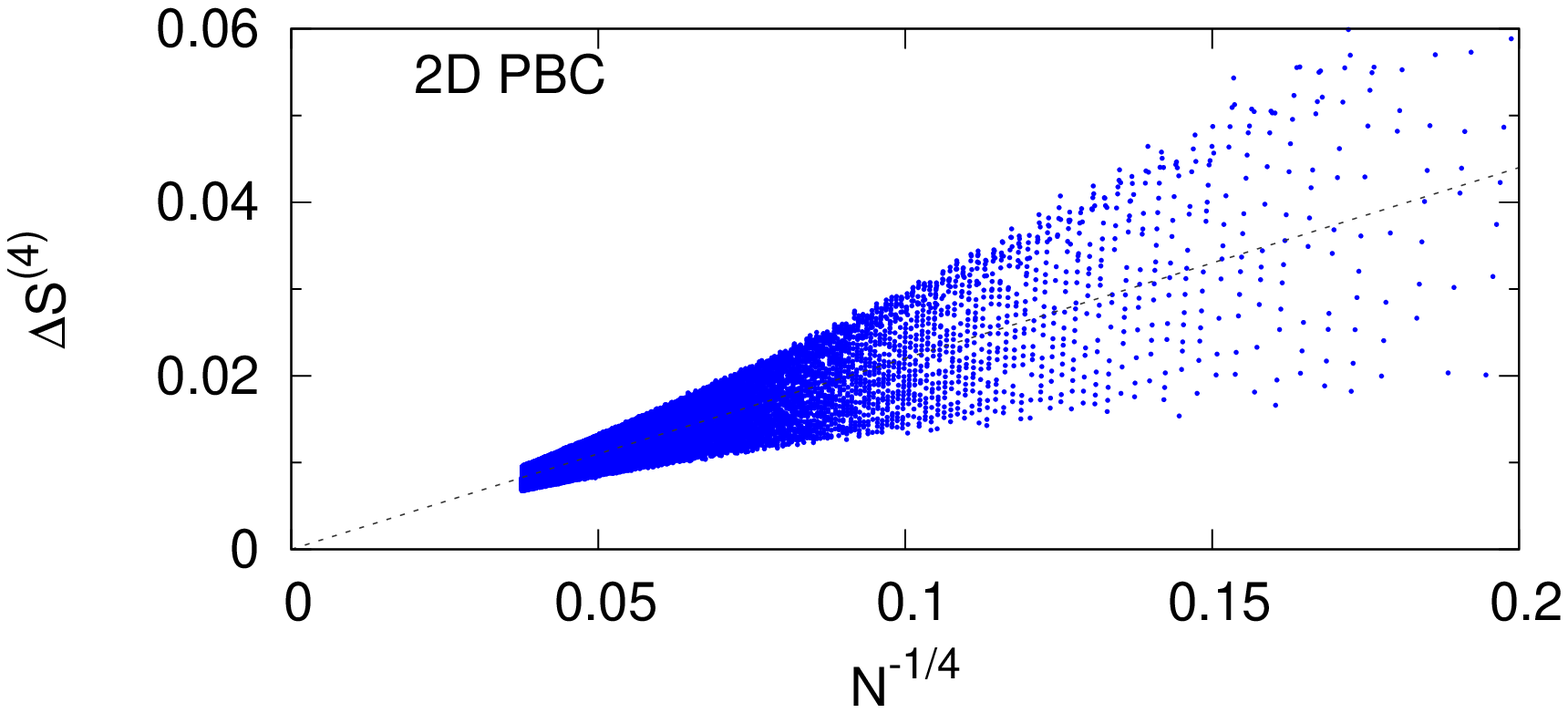}
\caption{(Color online) We plot the subtracted half-space vN and
  $\alpha = 4$ R\'enyi entropies defined in Eq.~(\ref{dsplot}), for a
  2D Fermi gas with PBC.  The powers of $N$ reported in abscissa are
  the expected power laws of the large-$N$ scaling
  corrections.  The lines are drawn to guide the eyes.  }
\label{2dShs}
\end{figure}

We now present finite-$N$ numerical results obtained by explicitly
computing the eigenvalues of the overlap matrix (\ref{aiodef}) and
plugging them into the formula (\ref{soma}) \cite{CMV-11, CMV-12-2}.
We compare their behavior with increasing $N$, up to $N=5\times 10^5$
and $N=3\times 10^7$ for 2D and 3D, respectively, with the exact
asymptotic behaviors derived above, and study the deviations, for
which we previously argued the power-law suppression. In particular,
we show that such corrections are characterized by peculiar
oscillations.

For this purpose, we show data for the half-space entanglement
entropies and particle variance after subtracting the asymptotic
behaviors computed in the previous subsections, i.e for the quantities
\begin{eqnarray}
\Delta O \equiv N^{-(d-1)/d}\left[O_{\rm
HS}(N)-O_{\rm HS, asy}\right],
\label{dsplot}
\end{eqnarray}
where the asymptotic formula for $S^{(\alpha)}$ and $V^{(2)}$ are
reported in Eqs.~(\ref{asysa}), (\ref{asyva}), (\ref{asysahs2}),
(\ref{asyv2hs2}), (\ref{sa3D}), and (\ref{asysahs33}), for 2D and 3D,
with PBC and OBC.  
The data for OBC refer to the half-space region defined in 
Eq.~(\ref{Astript}).

\begin{figure}[tbp]
\includegraphics[width=8.5cm]{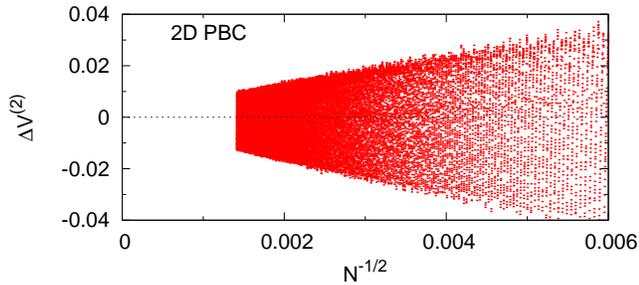}
\caption{(Color online) We plot the subtracted half-space particle
  variance defined as in Eq.~(\ref{dsplot}), for a 2D Fermi gas with
  PBC. }
\label{2dVhs}
\end{figure}

\begin{figure}[tpb]
\includegraphics[width=8.5cm,keepaspectratio=true]{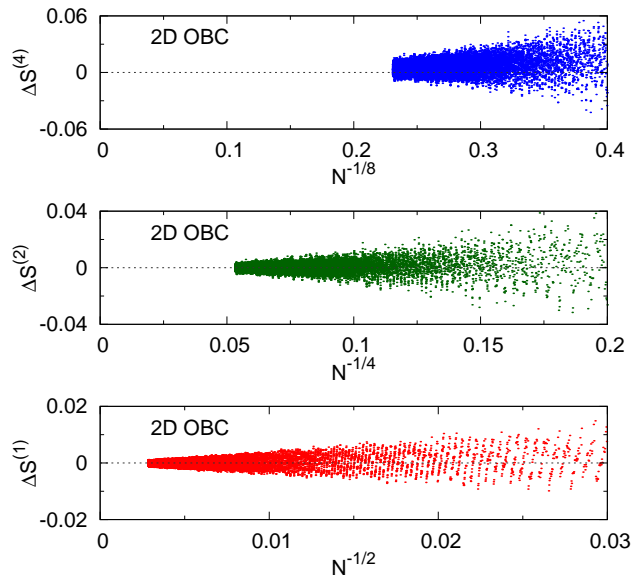}
\caption{(Color online) We plot the subtracted half-space vN and
  $\alpha = 2,\,4$ R\'enyi entropies defined in Eq.~(\ref{dsplot}),
  for a 2D Fermi gas with OBC.  The powers of $N$ reported in abscissa
  are the expected power laws of the large-$N$ scaling corrections.  }
\label{2dShsobc}
\end{figure}

\begin{figure}[tbp]
\includegraphics[width=8.5cm,keepaspectratio=true]{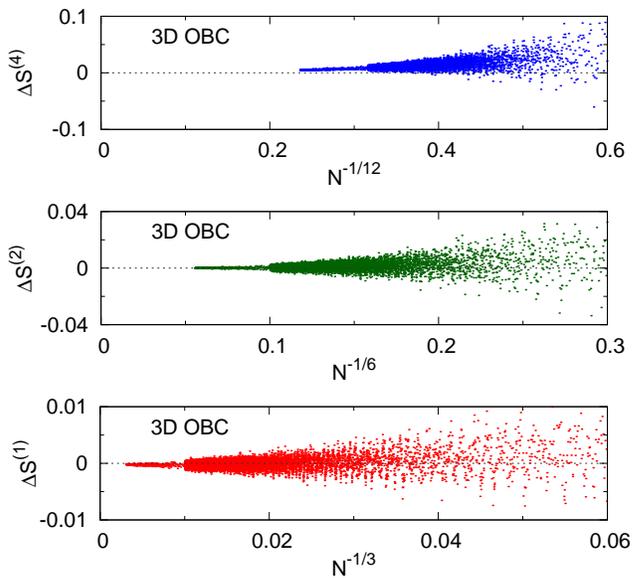}
\caption{(Color online) We plot the subtracted half-space vN and
  $\alpha = 2,\,4$ R\'enyi entropies defined in Eq.~(\ref{dsplot}),
  for a 3D Fermi gas with OBC.
The powers of $N$ reported in abscissa are the 
expected power laws of the large-$N$ scaling corrections.
}
\label{3dShs}
\end{figure}

Figs.~\ref{2dShs} and \ref{2dVhs} report results for the vN and
R\'enyi entanglement entropies and the particle variance,
respectively, for the half space of 2D systems with PBC.  The data
support the asymptotic behaviors obtained previously. In particular,
the subtracted quantities (\ref{dsplot}) appear to vanish with
increasing $N$ with the expected power laws (possibly multiplied by a
logarithmic factor), i.e. $N^{-\kappa}$ with $\kappa={\rm
min}[1/2,1/\alpha]$ for the entanglement entropies and $N^{-1/2}$ for
the particle variance.  Note however that such corrections are
characterized by wide oscillations.  Analogous results are obtained
for the half-space entanglement entropies in the case of OBC, as shown
by Figs.~\ref{2dShsobc} and \ref{3dShs}, respectively in 2D and 3D
systems.

\begin{figure}[t]
 \centering
\includegraphics[width=8.5cm,keepaspectratio=true]{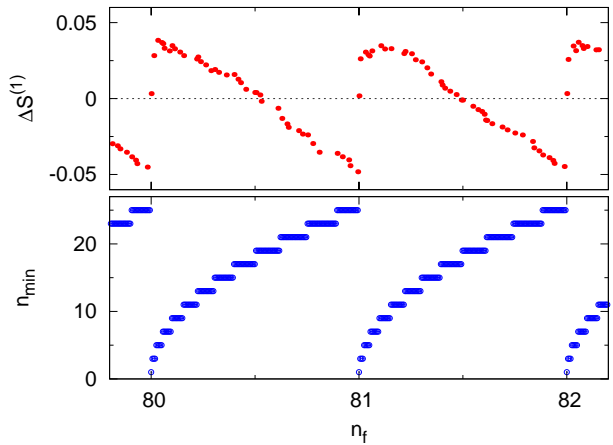}
\caption{(Color online) Top: we plot the quantity (\ref{dsplot})
for the half-space vN entanglement entropy of a 2D Fermi gas with PBC,
versus $n_f$. The oscillating corrections show a periodic
behavior with respect to $n_f$, with period $\Delta n_f = 1$. 
Bottom: for comparison, we also plot the size $n_{{\rm min}}$ of
  the smallest 1D block intervening in the calculation, which
presents similar oscillations.
}
 \label{fig:svn_blocks}
\end{figure}

The oscillations of 2D systems exhibit a distinctive structure, and
are essentially related to the way the Fermi sphere is filled as we
add more particles to the system.  In particular, it is interesting to
note that an oscillation of the vN entanglement entropy corresponds to
$n_f \to n_f+1$, cf. Eq.~(\ref{nedef}).
This is shown by the data of
Fig.~\ref{fig:svn_blocks} for the vN entanglement entropy of 2D
systems with PBC.

For any integer $n_f$, the smallest block contributing to the
calculation is of size $n_{\rm min} = 1$, and is obtained by setting
$n_1 = \pm n_f$ in the right hand side of Eq.~(\ref{oatrap}).  Note
that the size of the second smallest block, $n^{(2)}_{\rm min}$,
corresponding to $n_1 = \pm (n_f-1)$, grows with $n_f$ as
\begin{equation}
n^{(2)}_{\rm min} (n_f)= 2 \lfloor \sqrt{2 n_f -1} \rfloor 
                   \longrightarrow 2 \lfloor \sqrt{2n_f}\rfloor
                   \sim N^{1/4}.
\label{eq:n1min} 
\end{equation}
Therefore, except for the blocks of size one, which contribute with a
constant, the smallest blocks are asymptotically large, thus
justifying the use of asymptotic formulas in Eq.~(\ref{oatrap}).
However, if we release the requirement that $n_f$ be an integer,
$n_{\rm min}$ is in general greater than one, and Eq.~(\ref{eq:n1min})
provides an upper bound to it.  Following the quantity $n_{\rm min}$
from an integer $n_f$ to $n_f+1$, $n_{\rm min}$ starts by taking unit
value, then monotonically increases up to $n_{\rm min}^{(2)}(n_f +
1)$, finally dropping again to one (cf. bottom panel of
Fig.~\ref{fig:svn_blocks}).  A comparison of data for $n_{\rm min}(N)$
with the oscillating corrections to the asymptotic behavior of the
half-space vN entanglement entropy is shown in
Fig.~\ref{fig:svn_blocks}.  The connection between the two quantities
is evident, and can be heuristically explained by considering the
``specific entropy'' carried by each 1D block that contributes to the
sum Eq.~(\ref{oatrap}). According to Eq.~(\ref{FHres2pbc}), we can
define the specific entropy of a block $s(n) = S_{\rm 1D}(n)/n \sim
(\ln n)/n$, which is a function with an absolute maximum for $n\approx
0.3$.  The specific entropy is then monotonically decreasing for any
integer $n$.  Small blocks add few particles but lots of entropy, and
thus give a positive correction over the asymptotic formula
(\ref{sadetrapr}), which in the subtracted data plotted in
Fig.~\ref{fig:svn_blocks} approximately correspond to the line $y=0$.
On the other hand, larger blocks add many particles but relatively
little entropy, and hence explain the descending part of the
oscillation.  This reasoning only makes use of the geometric
characteristics of the Fermi sphere and of the leading behaviour of
the entanglement entropies for 1D gases, which is shared among all the
quantities we study in this paper.  Therefore the same explanation of
the oscillations apply to the particle variance as well.

In dimension higher than two, a similar analysis of the oscillations
would lead to a decomposition in terms of superpositions of many 1D
blocks at once, thus explaining the more complex structure of higher
order corrections to the entanglement entropy shown in
Fig.~\ref{3dShs}.  Note also that in the case of the R\'enyi entropies
($\alpha\ge 2$) and higher cumulants, as well as for strips of width
other than the half space, some oscillations already appear in the
correction to the asymptotic 1D behavior~\cite{CMV-11-2,CMV-12-2}, and
may result in further contributions and modulations of the oscillatory
corrections in 2D.

\section{Fermi gas on a disk}
\label{sphvol}

In order to understand how extended observables may depend on the
shape of the hard-wall trap, we consider a Fermi gas in a disk $D$ of
unit radius $R=1$ with hard-wall boundary conditions.  Its many-body
wave function is still given by Eq.~(\ref{fpsi}) with one-particle
eigenfunctions satisfing the boundary condition $\psi_k(R,\theta)=0$
in spherical coordinates.  The one-particle normalized eigenspectrum
is given by
\begin{eqnarray}
\psi_{nl}(r,\theta) = {1\over\sqrt{\pi}
J_{|l|+1}(k_{nl})}  J_{|l|}(k_{nl} r) e^{il\theta},
\label{psimnfp}
\end{eqnarray}
where $l\in {\mathbb Z}$ and $n \in {\mathbb N}$, $k_{nl}>0$ is the
$n^{\rm th}$ zero (excluding the origin) of the Bessel function
$J_{|l|}$, and the energy levels are $e_{nl} = {k_{nl}^2/2}$.

Let us consider a subsystem $A$ given by a disk of radius $r<R=1$. In
order to derive the leading large-$N$ behavior of their entanglement
entropy, we use the large-area asymptotic formula~\cite{CMV-12}
\begin{equation}
S^{(\alpha)}_A \approx  {1+\alpha^{-1}\over 6\sqrt{\pi}} \rho^{1/2} 
{\cal A} \ln {\cal A},
\label{SAa}
\end{equation}
where ${\cal A}$ is the area separating $A$ from the rest, and $\rho$
is the particle density.  This result was derived~\cite{CMV-12} by
turning the large-$N$ behavior of $\ell_1\times \ell_2$ subsystems in
square $L^2$ systems into the large-area behavior in the thermodynamic
limit keeping $\rho=N/V$ fixed.  Note that since we are now
considering a different geometry we cannot straighforwardly use the
large-$N$ results of rectangular subsystems in square
systems. However, we expect that, in the thermodynamic limit keeping
$\rho\equiv N/V$ fixed, the coefficient of the logarithmic correction
to the area law depends only on the particle density.  In other words,
when it is expressed in terms of the local particle density, it is
expected to be independent of the system shape.  Then we turn the
large-area asymptotic behavior (\ref{SAa}) into a large-$N$ asymptotic
behavior assuming that the limits can be interchanged.  In the case at
hand, replacing ${\cal A}=2\pi r$ and $N=\pi R^2\rho$ into
Eq.~(\ref{SAa}), we obtain
\begin{equation}
S^{(1)}(r) \approx {r\over 3} N^{1/2}\ln N
\label{sr}
\end{equation}
for the vN entropy.

\begin{figure}[tbp]
\includegraphics*[scale=\graphicscale]{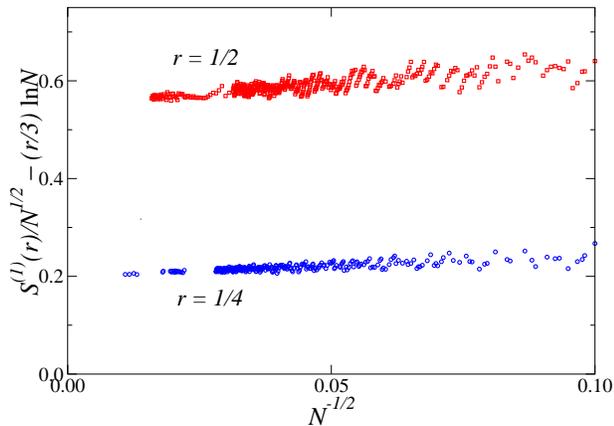}
%=% \vskip-5mm
\caption{ (Color online) The vN entanglement entropy of disks of
  radius $r=1/4$ and $r=1/2$ with respect to the rest, in Fermi gases
  constrained within a disk of of size $R=1$.  We plot
  $N^{-1/2}S^{(1)}(r) - {r\over 3}\ln N$ vs $N^{-1/2}$.  }
\label{s1}
\end{figure}

We check this asymptotic behavior by finite-$N$ computations based on
the overlap matrix (\ref{aiodef}) using the one-particle eigenspectrum
(\ref{psimnfp}).  Some numerical results are reported in Fig.~\ref{s1}
for $r=1/4$ and $r=1/2$.  They clearly support Eq.~(\ref{sr}), with
the same pattern of corrections found in the systems considered in the
previous section, i.e.
\begin{equation}
S^{(1)}(r) = {r\over 3} N^{1/2}(\ln N + a_0) + O(N^0),
\label{sr2}
\end{equation}
and show again oscillations, which appear to increase with increasing
$r$.

\section{Fermi gases released from a trap}
\label{reltrap}

In this section we consider the free expansion of the Fermi gas after
the instantaneous drop of the hard-wall trap $T=[-L/2,L/2]^d$,
described by the many-body wave function reported in
Sec.~\ref{freexp}.  In the following we set $L/2=1$; the dependence on
the trap size can be easily recovered by a dimensional analysis.  We
determine the time evolution of entanglement entropies and particle
fluctuations for a large number of particles.

\subsection{The large-$N$ limit}
\label{lnft}

The equal-time two-point function can be written as
\begin{eqnarray}
&&C({\bf x},{\bf y},t) = \label{tpfe}\\ 
&&
\sum_{k=1}^N \int_T
d^d{\rm z}_1\,d^d{\rm z}_2 \,
{\cal P}({\bf x},t;{\bf z}_1,0)
{\cal P}({\bf y},t;{\bf z}_2,0) 
\,\phi_k({\bf z}_1) \phi_k({\bf z}_2) 
\nonumber
\end{eqnarray}
where $T={[-1,1]^d}$.
The large-$N$ limit of the equal-time two-point function can be
obtained using the completeness relation~\cite{Landau-book}
\begin{equation}
\sum_{k=1}^\infty \phi_k({\bf x}) \phi_k({\bf y}) = \delta({\bf x}-{\bf y})
\label{comprel}
\end{equation}
of the spectrum of the one-particle Hamiltonian at $t=0$.  We obtain
the large-$N$ limit
\begin{eqnarray}
C_\infty({\bf x},{\bf y},t) = 
\prod_{i=1}^d 
{\sin[(y_i-x_i)/t] \over \pi (y_i-x_i)}\, e^{i(x_i^2-y_i^2)/(2t)}.
\label{tpfln}
\end{eqnarray}
This implies that the large-$N$ limit of the time evolution of the
particle density is simply
\begin{equation}
\rho_\infty({\bf x},t) = C_\infty({\bf x},{\bf x},t) = {1\over (\pi t)^d},
\label{rhoxtis}
\end{equation}
which is independent of the position. Of course, this regime is approached
nonuniformly with respect to the spatial coordinate, but it is rapidly
reached around the central region of the size of the original trap,
i.e. $|{\bf x}|\lesssim 1$.  The density-density correlation function
can be obtained using Eq.~(\ref{gnbost}).

In order to compute quantities that depend only on the eigenvalues of
the restriction $C_A$ of $C$ within an extended region $A$, such as
the entanglement entropies, we can further simplify Eq.~(\ref{tpfln})
by dropping the phases, retaining only
\begin{eqnarray}
&&{\mathbb C}_A({\bf x},{\bf y},t) = \prod_{i=1}^d {\mathbb
Y}(x_i,y_i,t),\label{hatc}\\ && {\mathbb Y}(x,y,t)={\sin[(y-x)/t]
\over \pi (y-x)} .
\label{ffded}
\end{eqnarray}

Let us now consider an extended cubic-like subsystem
\begin{equation}
A=[-s,s]^d,
\label{assdef}
\end{equation}
with $s$ of the size of the original trap.  The large-$N$ particle
number $n_\infty(t)$ and cumulants $V_\infty^{(m)}(t)$ of the subsystem
$A$ can be written in terms of the 1D traces
\begin{equation}
{\cal T}_k \equiv {\rm Tr}_{[-s,s]}\,{\mathbb Y}^k,
\label{tkdef}
\end{equation}
so that
${\rm Tr} \,{\mathbb C}_A^k = {\cal T}_k^d$.
Indeed we have
\begin{eqnarray}
n_\infty(t) = {\cal T}_1^d = \left( {2s\over \pi t}\right)^d,
\label{ntln}
\end{eqnarray}
and
\begin{eqnarray}
V_\infty^{(2)}(t) &=&
{\rm Tr} \,{\mathbb C}_A ( 1 - {\mathbb C}_A) =  
{\cal T}_1^d - {\cal T}_2^d,
\label{v2infty} \\
V_\infty^{(3)}(t) &=& 
{\rm Tr} \,({\mathbb C}_A - 3 {\mathbb C}_A^2 + 2 
{\mathbb C}_A^3) =  
{\cal T}_1^d - 3 {\cal T}_2^d + 2 {\cal T}_3^d,
\nonumber\\
V_\infty^{(4)}(t) &=& {\rm Tr} [{\mathbb C}_A - 
7 {\mathbb C}_A^2 + 12 {\mathbb C}_A^3 - 6 {\mathbb C}_A^4] =
\nonumber\\
&=& {\cal T}_1^d - 7 {\cal T}_2^d + 12 {\cal T}_3^d
- 6 {\cal T}_4^d,
\nonumber
\end{eqnarray}
etc. In particular, for the space region $A=[-1,1]^d$, i.e. $s=1$,
which coincides with the original trap, we obtain
 \begin{eqnarray}
&&{\cal T}_1 = {2 \over \pi t}, \label{calt1} \\
&&{\cal T}_2 = 
\int_{-1}^1 dx \,dy \,{\mathbb Y}(x,y)^2=
\label{tranex}\\
&&= {4 {\rm Si}(4/t) 
-t[1 + \gamma_E + {\rm ln}(4/t)-{\rm cos}(4/t)- {\rm Ci}(4/t)]
\over \pi^2 t} 
\nonumber
\end{eqnarray}
where Ci and Si are the cosine and sine integral functions.  The
small-$t$ asymptotic expansion of the particle variance can be easily
derived, obtaining
\begin{eqnarray}
V_{\infty,{\rm asy}}^{(2)}(t) &=&  {2^{d-1} d\over \pi^{d+1}} t^{1-d}  
\left[ \ln(1/t) + 1+\gamma_E + 2 \ln 2 \right] 
\nonumber \\
&+& O\left[ t^{2-d} \ln^2(1/t)\right].
\label{lnv2asmallt}
\end{eqnarray}

Note that the above calculations can be easily extended to anisotropic
rectangular-like traps and/or subsystems, although they become more
cumbersome.

\subsection{The small-$t$ behavior of particle cumulants and
bipartite entanglement entropies}
\label{smallt}

\begin{figure}[tbp]
\includegraphics*[scale=\graphicscale]{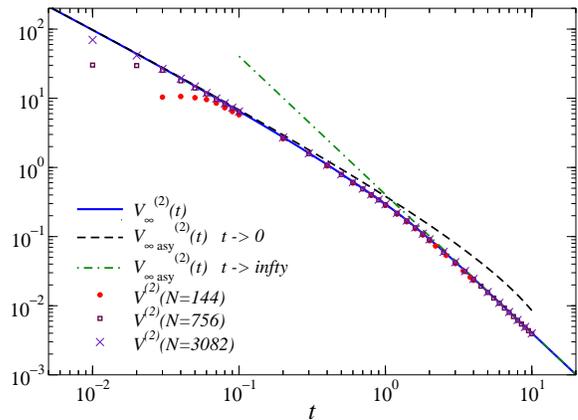}
\caption{(Color online) For 2D Fermi gases released by a hard-wall
  trap $[-1,1]^2$ and a spatial region corresponding to the initial
  trap, we compare the large-$N$ limit $V_\infty^{(2)}(t)$ of the
  particle variance, obtained from Eqs.~(\ref{v2infty}),
  (\ref{calt1}), (\ref{tranex}), with its small-$t$ and large-$t$
  asymptotic behaviors, Eqs.~(\ref{vasyd}) and (\ref{vma1})
  respectively.  Moreover, we also show finite-$N$ data which approach
  the large-$N$ results, nonuniformly with decreasing $t$.}
\label{v2t}
\end{figure}

The Eqs.~(\ref{v2infty}) allow us to relate the small-$t$
asymptotic behaviors of the particle cumulants to those obtained in 1D
Fermi gases released from a 1D trap $[-1,1]$ and associated with the
interval $[-s,s]$, which are~\cite{V-12-2}
\begin{eqnarray}
&&{\cal V}_{\rm asy}^{(2)}(t) \approx 
{1\over \pi^2}\left[ \ln(1/t) + \ln 2s  + v_2\right],
\label{asy1dv2}\\
&&{\cal V}_{\rm asy}^{(2k+1)}(t) \approx 0  \qquad  (k\ge 1), 
\label{lnpred3a}\\
&&{\cal V}_{\rm asy}^{(2k)}(t) \approx   v_{2k}   \qquad\, (k\ge 2), 
\label{lnpred4a}
\end{eqnarray}
where $v_2$, $v_4$, etc. are the same constants appearing in in
Eqs.~(\ref{v21d}) and (\ref{v461d}).  Indeed, these asymptotic
expressions allow us to derive the small-$t$ behaviors of ${\cal
T}_k$, cf. Eq.~(\ref{tkdef}).  By inserting them in
Eq.~(\ref{v2infty}), we obtain the $d$-dimensional asymptotic relation
\begin{equation}
V^{(m)}_{\infty,{\rm asy}}(t) = 
d \left( {2s\over \pi t}\right)^{d-1}
\times {\cal V}_{\rm asy}^{(m)}(t).
 \label{vasyd}
\end{equation} 
Corrections to this asymptotic behaviors are suppressed by powers of
$t$ with respect to the leading term (with possible multiplicative
logarithms).  
In the case of the particle variance the asymptotic behavior
(\ref{vasyd}) coincides with Eq.~(\ref{lnv2asmallt}).  The comparison
with the exact large-$N$ limit $V_\infty^{(2)}(t)$, see
Fig.~\ref{v2t}, shows that for 2D systems Eq.~(\ref{vasyd}) provides a
good approximation when $t\lesssim 0.2$.

Therefore, Eq.~(\ref{vasyd}) shows that,
like 1D systems~\cite{V-12-2}, only the particle variance
has a multiplicative logarithmic correction to the leading power law
in the small-$t$ limit, behaving as $V^{(2)}_{\infty}(t)\sim
t^{1-d}\ln(1/t)$, while higher even cumulants behave simply as
$V^{(2k)}\sim t^{1-d}$, and odd cumulants are relatively suppressed.

In order to derive the small-$t$ asymptotic behaviors of the
entanglement entropies, we may use their relation (\ref{sv2ia}) with
the particle cumulants. Since only the particle variance has the
leading multiplicative logarithm, we obtain the asymptotic small-$t$
relation
\begin{equation} 
{S^{(\alpha)}_{\infty}(t)\over V^{(2)}_{\infty}(t)} \approx 
\pi^2 c_\alpha , \label{anydt} 
\end{equation}
where corrections are only logarithmically suppressed.

It is worth mentioning that the same result can also be obtained
following the procedure employed in Ref.~\onlinecite{V-12-2} (see its
Sec.~IIIB), based on a discretization of the two-point function
(\ref{ffded}), which maps it to a lattice free-fermion model without
boundary, at equilibrium in the thermodynamic limit and with a cubic
Fermi surface with $k_F\sim 1/t$.  Then, the asymptotic small-$t$
behavior of the entanglement entropy can be evaluated using the Widom
conjecture~\cite{Widom-81}, using e.g.  the formulas of
Ref.~\cite{CMV-12-2}.

Actually, we can also determine the next-to-leading corrections of the
entanglement entropies, by noting that the $O(t^{1-d})$ asymptotic
behavior of $d$-dimensional cumulants is just obtained by multiplying
the same factor times the 1D results, cf. Eq.~(\ref{vasyd}). Thus,
using Eq.~(\ref{sv2ia}), we conjecture that the same result
extends to the entanglement entropies, i.e.
\begin{equation}
S^{(\alpha)}_{\infty,{\rm asy}}(t) 
\approx d \left({2s\over \pi t}\right)^{d-1} \times {\cal S}^{(\alpha)}(t)
\label{sasyt}
\end{equation}
where ${\cal S}^{(\alpha)}$ is the asymptotic small-$t$ behavior 
of 1D systems for the interval $[-s,s]$, i.e.~\cite{V-12-2}
\begin{eqnarray}
{\cal S}^{(\alpha)}(t) = c_\alpha
\left[\ln(1/t) + \ln 2s + b_\alpha\right] 
\label{lnpred1a}
\end{eqnarray}
where $c_\alpha$ and $b_\alpha$ are given in Eqs.~(\ref{calpha}) and
(\ref{ba}).  In the small-$t$ regime corrections to Eq.~(\ref{sasyt})
are expected to be relatively suppressed by powers of $t$.

\subsection{The large-time behavior of entanglement entropies
and particle fluctuations}
\label{ltbeh}

We now compute the large-$t$ behaviors of particle fluctuations and
entanglement entropies of a space region $[-s,s]^d$.  They can be
inferred by looking at the behavior of the eigenvalues of the overlap
matrix, observing that only one eigenvalue dominates at large time.
Indeed, using Eq.~(\ref{acrel}) and the large-time behavior of
${\mathbb C}_{A}$, we have
\begin{equation}
{\rm Tr}\,{\mathbb A}^k = {\rm Tr}\,{\mathbb C}_{A}^k \sim 
\left( {2s\over \pi t}\right)^{dk}
\label{asyacrel}
\end{equation}
for any integer $k$.
This implies that the large-$t$ regime of the overlap matrix is
characterized by only one nonzero eigenvalue $a_1$ for any $N$, given
by
\begin{eqnarray}
a_1 \approx \left( {2s\over \pi t}\right)^d.
\label{onlyeig}
\end{eqnarray}
The other eigenvalues get rapidly suppressed in the large-$t$ limit.

The largest eigenvalue $a_1$ determines the asymptotic large-$t$
behaviors of the particle number, particle fluctuations and
entanglement entropies:
\begin{eqnarray}
&&n_A(t)\approx  V_A^{(m)} \approx a_1  \quad 
{\rm for\;any}\;\;m,\label{vma1}\\
&&
S_A^{(\alpha)} \approx {\alpha\over \alpha-1} a_1,\quad
S_A^{(1)} \approx a_1 (1 - \ln a_1).
\label{alllt}
\end{eqnarray}

As shown in Fig.~\ref{v2t}, in 2D systems the large-$t$ behavior
(\ref{vma1}) provides a good approximation of the exact large-$N$
limit (\ref{v2infty}) of the particle variance for $t\gtrsim 2$.

\subsection{Convergence to the large-$N$ limit}
\label{finiten}

\begin{figure}[tbp]
\includegraphics*[scale=\graphicscale]{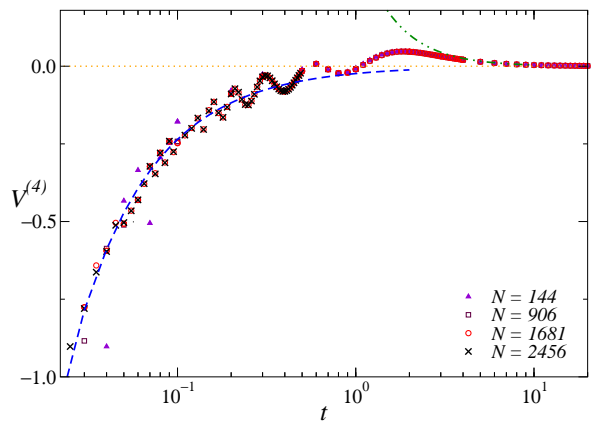}
%=% \vskip-5mm
\caption{(Color online) Finite-$N$ results for the quartic
cumulant $V^{(4)}$, compared with the large-$N$ results 
in the small-$t$ regime (dashed line),
  cf. Eq.~(\ref{vasyd}), and in the large-$t$ regime (dot-dashed line),
  cf. Eq.~(\ref{vma1}).  }
\label{v4t}
\end{figure}

\begin{figure}[tbp]
\includegraphics*[scale=\graphicscale]{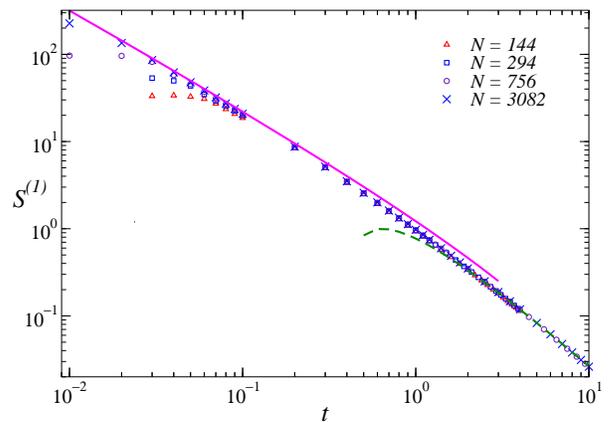}
%=% \vskip-5mm
\caption{(Color online) For 2D Fermi gases released by a hard-wall
  trap $[-1,1]^2$ and the vN entanglement entropy of a spatial region
  corresponding to the initial trap, we compare finite-$N$ data with
  the analytical large-$N$ results in the small-$t$ regime (full
  line), cf. Eq.~(\ref{sasyt}), and in the large-$t$ regime (dashed
  line), cf. Eq.~(\ref{alllt}).  }
\label{s1t}
\end{figure}

We want to check the actual convergence of the particle fluctuations
and entanglement entropies of systems with $N$ particles to the
large-$N$ behaviors derived in the preceding subsections, which is
generally expected to be chacterized by asymptotic $O(N^{-1/d})$
corrections.  For this purpose, we compute them at finite particle
number $N$ using the overlap matrix, see Sec.~\ref{mbwaf}.  We
numerically compute its eigenvalues at fixed $t$, and then obtain the
particle cumulants and the entanglement entropies, see
Ref.~\cite{V-12-2} for more details.

\begin{figure}[tbp]
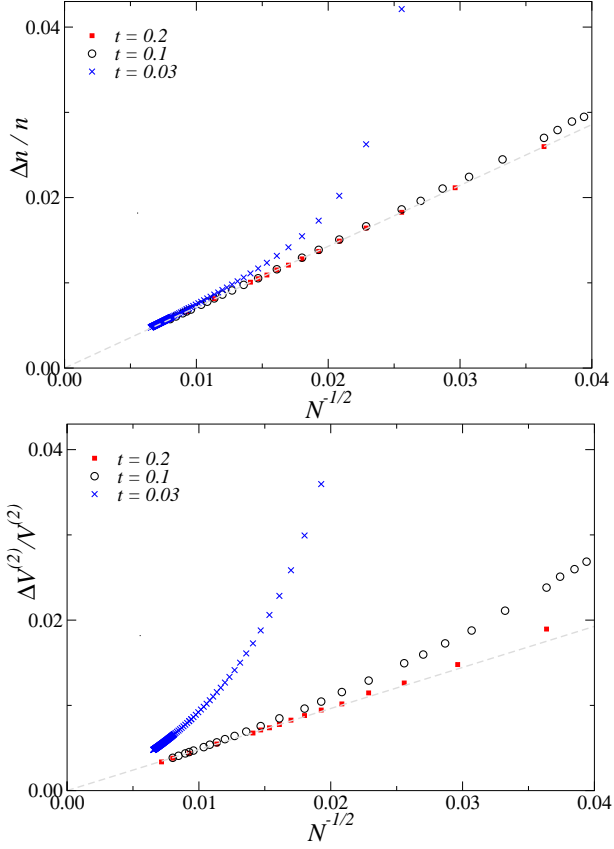

\includegraphics*[scale=\graphicscale]{fig11a.eps}
\includegraphics*[scale=\graphicscale]{fig11b.eps}
\caption{ (Color online) In order to show the convergence to the
  large-$N$ limit of the particle number $n(t)$ and variance
  $V^{(2)}(t)$ of the space region $A=T=[-1,1]^2$, we plot the relative
  differences $\Delta n(t)/n_\infty(t)$ and $\Delta
  V^{(2)}(t)/V^{(2)}_\infty(t)$, where $\Delta n(t) = n_\infty(t) -
  n(t;N)$ and $\Delta V^{(2)}(t) = V^{(2)}_\infty(t) - V^{(2)}(t;N)$.
  }
\label{nv2fixedt}
\end{figure}

\begin{figure}[tbp]
\includegraphics*[scale=\graphicscale]{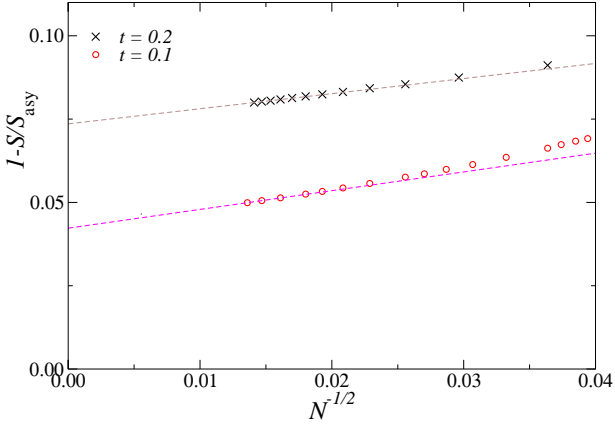}
\caption{ (Color online) Data for the vN entanglement entropy of the
space region $[-1,1]^2$ corresponding to the initial trap, at $t=0.1$
and $t=0.2$, and several values of $N$.  We plot
$1-S^{(1)}(t)/S^{(1)}_{\infty,{\rm asy}}$ where $S^{(1)}_{\infty,{\rm
asy}}$ is the asymptotic (nonexact) formula (\ref{sasyt}).  The dashed
lines are linear extrapolations of the data for the largest values of
$N$.  The extrapolated nonzero values are consistent with the expected
$O(t)$ corrections 
to the leading small-$t$ terms (\ref{sasyt}).  }
\label{s1fixedt}
\end{figure}

We report results for 2D Fermi gases, and subsystems corresponding to
the initial square trap.  In Fig.~\ref{v2t} we show data for the
particle variance. They approach the large-$N$ limit analytically
evaluated using Eqs.~(\ref{v2infty}), (\ref{calt1}) and
(\ref{tranex}), although nonuniformly with decreasing $t$.
Fig.~\ref{v4t} shows analogous data for the quartic cumulant
$V^{(4)}$, which is characterized by an undulatory large-$N$ limit.
In Fig.~\ref{s1t} we show results for the vN entanglement entropy:
again the data approach a large-$N$ limit, which is well approximated
by the large-$N$ analytical formulas in the small-$t$ and large-$t$
regime, cf. Eq.~(\ref{sasyt}) and (\ref{alllt}), for $t\lesssim 0.2$
and $t\gtrsim 2$ respectively.  The large-$N$ convergence is more
clearly demonstrated in Fig.~\ref{nv2fixedt}, where we show data of
the particle number and particle variance versus $N^{-1/2}$ for some
small and fixed values of $t$.  Analogous results are shown in
Fig.~\ref{s1fixedt} for the vN entanglement entropy.

As expected the convergence to the large-$N$ limit at fixed $t$ is
characterized by leading $O(N^{-1/2})$ corrections, but it does not
appear uniform when $t\to 0$.  This may suggest some other nontrivial
scaling behaviors at small $t$, which get hidden (pushed toward the
$t=0$ axis) by the large-$N$ limit keeping $t$ fixed, as found in 1D
systems~\cite{V-12-2}. This point may deserve further investigation,
but we have not pursued it.

\section{Conclusions}
\label{conclusions}

We study the quantum correlations of 2D and 3D Fermi gases, as
described by extended observables such as entanglement entropies and
particle fluctuations associated with finite space regions. We mainly
consider two physical situations.  The Fermi gas of $N$ particles is
confined within a limited space region by a hard-wall trap and is at
equilibrium at $T=0$, i.e. in its ground state. Then we consider its
nonequilibrium unitary dynamics after the Fermi gas is released from
the trap, due to an instantaneous dropping of the trap.  This is one
of the simplest examples of nonequilibrium unitary evolution of
many-body systems, which may provide interesting general information
on the nonequilibrium entanglement properties, also in dimensions
larger than one.

Within the ground state of 2D and 3D Fermi gases in cubic-shaped
hard-wall traps, we compute the leading and next-to-leading terms of
the asymptotic large-$N$ behaviors of entanglement entropies and
particle fluctuations of strip-like subsystems. Their simple geometry
allows us to significantly extend earlier calculations based on the
so-called Widom conjecture, which were limited to the leading term
without providing any information on the subleading terms.  For
example, in the case of a 3D Fermi gas in a cubic box of size $L=1$
with hard-wall boundary conditions, the half-space vN entanglement
entropy, i.e.  associated with the space region $[0,L/2]\times L^2$
(see Fig.~\ref{fig:subsys_diagram}b), behaves as
\begin{eqnarray}
S^{(1)}_H &=& 
{\pi^{1/3} \over 2^{7/3}3^{4/3}}
N^{2/3} \Bigl[ \ln N + 3b_1 +
\label{s1co}\\
&+&  \ln(24/\pi)
 - 3/2 + O(N^{-1/3})\Bigr]
\nonumber
\end{eqnarray}
where the constant $b_1$ comes from 1D calculations,
cf. Eq.~(\ref{ba}), and the suppressed corrections are also
characterized by peculiar oscillations, discussed in
Sec.~\ref{corrections}.

We then consider the nonequilibrium unitary evolution of the Fermi
gases after instantaneously dropping the hard-wall trap.  We
investigate the entanglement properties during the expansion, by
computing the vN and R\'enyi entanglement entropies and the particle
cumulants of extended space regions in proximity to the initial trap.
Like 1D systems~\cite{V-12-2}, in the large-$N$ limit the equal-time
two-point function assumes a relatively simple form, given by
Eq.~(\ref{tpfln}), which can be further simplified when considering
only the eigenvalues of its space restrictions, effectively reducing
it to
\begin{equation}
{\mathbb C}({\bf x},{\bf y},t)=\prod_i {\sin[(y_i-x_i)/t]\over \pi(y_i-x_i)} .
\label{cco}
\end{equation}
This allows us to derive some exact results on the particle cumulants,
in particular the particle variance, and the small-$t$ asymptotic
behavior of the entanglement entropies of space regions in proximity
to the original trap.  For example, the vN entanglement entropy of the
subsystem $[-1,1]^d$, coinciding with the original trap, behaves as
(we set $L=2$ for the size of the trap, cf. Eq.~(\ref{trapreg}), which
implies that time is measured in units of $m (L/2)^2/\hslash$)
\begin{eqnarray}
S^{(1)}_{\infty}(t) 
= {2^{d-1}d\over 3 \pi^{d-1}}\,  
t^{1-d} \left[ \ln(1/t) + \ln 2 + b_1 + O(t)\right],
\label{s1cot}
\end{eqnarray}
at small $t$, and
\begin{equation}
S^{(1)}_\infty(t) \approx d \left( {2\over \pi t}\right)^d \ln t
\label{s1ltco}
\end{equation}
in the large-$t$ limit.

These results are obtained in the large-$N$ limit keeping $t$ fixed,
which turns out not to be uniform when $t\to 0$. However, the analysis
of numerical results at finite $N$, obtained using the method based on
the overlap matrix, shows that the approach to the large-$N$ limit is
quite rapid, for example in 2D a few hundred particles are already
sufficient to show the main features of the large-$N$ time dependence.

We also note that, analogously to 1D systems, also in higher
dimensions the small-$t$ dependence of the entanglement entropy and
particle fluctuations show some analogies with the asymptotic
behaviors at equilibrium.  For example, in the small-$t$ regime
$S^{(1)}_A \approx (\pi^2/3) V^{(2)}_A$, which has been shown to be
valid for the ground state of a large number of noninteracting Fermi
particles, in any dimensions and for any subsystems, in homogeneous
and inhomogeneous conditions~\cite{CMV-12-2,V-12}.  Moreover, the
small-$t$ dependence (\ref{s1cot}) shows multiplicative logarithmic
corrections to the leading power law, similarly to the area-law
violations (\ref{carelaw}).

The analogies between the small-$t$ and equilibrium asymptotic
behaviors are essentially connected with the form of the equal-time
two-point function during the free expansion from a hard-wall trap.
Indeed, as noted in Sec.~\ref{smallt} and also discussed in
Ref.~\cite{V-12} within 1D systems, ${\mathbb C}({\bf x},{\bf y},t)$
corresponds to an appropriate continuum limit of the two-point
function of a lattice free-fermion model without boundaries, at
equilibrium in the thermodynamic limit and with a cubic Fermi surface
with $k_F\sim 1/t$.

An important remark is that some features of the nonequilibrium free
expansion from hard-wall traps are closely related to the initial
(almost homogenous) conditions. For example, Fermi gases starting from
harmonic traps, realized using external space-dependent potentials,
show other peculiar nonequilibrium entanglement
properties~\cite{V-12-2}, such as that the entanglement entropies can
be expressed as a global time-dependent rescaling of the space
dependence of the initial equilibrium entanglement entropy,
in any dimension.  

Within particle systems released from hard-wall traps, further
interesting issues may concern the universality of the results
reported in this paper, in particular the universality of the
small-time asymptotic behaviors of the entanglement entropies and
particle fluctuations, whether they are shared with other many-body
systems, and their stability against particle interactions.

\end{document}